\documentclass[aps,prd,preprint,nofootinbib]{revtex4-1}

\usepackage{amsmath,amssymb,amsfonts}
\usepackage{bm}
\usepackage{physics}
\usepackage{graphicx}
\usepackage{xcolor}
\usepackage{slashed}
\usepackage[most]{tcolorbox}
\begin{document}

\title{Hamiltonian Renormalization Group and Boundary Conditions in Gauge Theories}

\author{J. Gamboa}
\affiliation{Departamento de F\'isica, Universidad de Santiago de Chile, Santiago 9170020, Chile }

\begin{abstract}
We formulate the renormalization group as a first-order Hamiltonian
variational problem in an extended phase space of running couplings,
renormalized quantities and their conjugate variables. We distinguish
boundary terms required by the variational principle from generating
functions of the canonical RG map. In particular, the mixed generating
function exists even when the bulk action vanishes on shell. We apply
this construction to the Born--Oppenheimer--Berry description of
infrared QCD, distinguishing the infrared dressing scale from the
conventional Wilsonian scale. In the scale-covariant effective regime,
a stationary coupling coexists with nontrivial anomalous scaling and
mixing of collective quark--gluon and gluon--gluon excitations. We
construct the canonical transport of these branches and its endpoint
generating function, and derive the transport of the response to
infrared spectral data. The resulting dispersion relations distinguish
soft, finite-energy and energetically suppressed long-wavelength
branches. The formulation separates the information determined by local
RG equations from the physical matching data required to specify the
infrared effective theory.
\end{abstract}

\maketitle

\section{Introduction}

The renormalization group (RG) provides a fundamental framework
for understanding the scale dependence of quantum field theories
in both perturbative and nonperturbative regimes
\cite{Callan1970,Symanzik1970,WilsonKogut1974,Polchinski1984}.
Its conventional formulation describes the evolution of couplings
and renormalized quantities under changes of scale, while ensuring
the independence of physical observables from the arbitrary
renormalization scale. In gauge theories such as QCD, the RG plays
a particularly important role in relating the perturbative
ultraviolet regime to the nonperturbative dynamics that
characterizes the infrared sector.

The renormalization-group evolution can be formulated in terms
of the parameter
\begin{equation}
t=\ln\mu ,
\label{rgt}
\end{equation}
so that the flow equations take the form of first-order
differential equations in $t$. Conventionally, these equations
are integrated by specifying the renormalized quantities at
a reference scale. A different perspective emerges, however,
when the ultraviolet and infrared endpoints are considered
simultaneously. In this case, the RG can be formulated as a
boundary-value problem, raising the question of which physical
data characterize its endpoints and how these data enter an
appropriate variational principle.

The distinction between local evolution and boundary data
becomes particularly relevant in the deep infrared regime of
QCD. In the physical picture considered here, this regime is
characterized by collective degrees of freedom whose dynamics
incorporates an emergent geometric Berry contribution.
The corresponding infrared scale is associated with the
dressing of collective excitations and must be distinguished
from the conventional Wilsonian renormalization scale.
Consequently, the infrared evolution involves physical
information that is not necessarily contained in the
perturbative beta functions or anomalous dimensions of
microscopic QCD.

The purpose of this paper is to develop a Hamiltonian
variational formulation of the RG in which the boundary
conditions are treated explicitly. By introducing an extended
phase space containing the running couplings, the renormalized
quantities and their conjugate variables, we identify the
canonical structure of the RG evolution and construct the
generating functions associated with its endpoint data.
A central aspect of this formulation is the distinction
between the surface terms required by the variational
principle and the generating functions that characterize
the canonical RG map.

We apply this construction to the Born--Oppenheimer--Berry
description of infrared QCD, in which quark--gluon and
gluon--gluon collective excitations constitute two coupled
infrared channels. Within the scale-covariant effective
regime, the stationary behavior of the infrared coupling
coexists with a nontrivial evolution of the collective
dressing functions. Their canonical transport is governed
by the eigenvalues of an effective anomalous-dimension
matrix, which also determine the long-wavelength behavior
of the corresponding excitation energies. This provides
a common framework for discussing the geometric structure
of infrared QCD, its scaling properties and the physical
information associated with the RG boundary conditions.

The role of boundary conditions in first-order dynamics
can be illustrated by a simple mechanical example.
Consider the evolution of a spin in the presence of an
external magnetic field,
\begin{equation}
\frac{d{\bf S}}{dt}
=
\kappa\,{\bf S}\times{\bf B}.
\label{spineq}
\end{equation}
This dynamics admits a representation in terms of Grassmann
variables $\theta_i$ through the action
\cite{Berezin:1976eg,Galvao:1980cu}
\begin{equation}
S
=
\int_{t_1}^{t_2}dt
\left[
\frac{i}{2}\theta_i\dot{\theta}_i
+
\frac{i\kappa}{2}
\epsilon_{ijk}\theta_i\theta_j B_k
\right]
+
S_b ,
\label{spinaction}
\end{equation}
where
\begin{equation}
\{\theta_i,\theta_j\}=0 ,
\label{grassmann}
\end{equation}
and $S_b$ denotes a surface term to be specified below.
The corresponding Hamiltonian is
\begin{equation}
H
=
-\frac{i\kappa}{2}
\epsilon_{ijk}\theta_i\theta_j B_k .
\label{spinham}
\end{equation}
Upon quantization, the Grassmann variables are represented by
\begin{equation}
\theta_i
\longrightarrow
\frac{1}{\sqrt{2}}\sigma_i ,
\label{grassquant}
\end{equation}
and the Hamiltonian becomes
\begin{equation}
H_b
=
\frac{\kappa}{2}\,
{\boldsymbol\sigma}\cdot{\bf B},
\label{pauliham}
\end{equation}
which is the Pauli Hamiltonian, up to the convention adopted
for the sign and normalization of the magnetic moment.

For our purposes, the relevant feature of this system is
the relation between its first-order dynamics and its
boundary data. Varying the bulk part of
Eq.~(\ref{spinaction}), one obtains the equations of motion
together with the endpoint contribution
\begin{equation}
\left.\delta S_{\rm bulk}\right|_{\partial}
=
\frac{i}{2}
\left[
\theta_i(t_2)\delta\theta_i(t_2)
-
\theta_i(t_1)\delta\theta_i(t_1)
\right].
\label{spinbulkvar}
\end{equation}
The local equations of motion therefore determine the
dynamics in the interior of the interval, whereas the
endpoint contribution must be considered separately in
order to complete the variational problem.

In the functional representation of a fermionic trace,
the Grassmann variables satisfy antiperiodic boundary
conditions,
\begin{equation}
\theta_i(t_2)=-\theta_i(t_1).
\label{antiperiodic}
\end{equation}
The physical origin of this condition lies in the fermionic
character of the functional trace rather than in the local
equations of motion. Indeed, the classical spin-precession
equation (\ref{spineq}), expressed solely in terms of the
bosonic spin variables ${\bf S}$, contains no information
about the fermionic antiperiodicity.

The compatibility of the variational principle with these
boundary conditions can be made explicit by introducing
the surface term
\begin{equation}
S_b
=
\frac{i}{2}
\theta_i(t_1)\theta_i(t_2).
\label{spinbdry}
\end{equation}
Its variation is
\begin{equation}
\delta S_b
=
\frac{i}{2}
\left[
\delta\theta_i(t_1)\theta_i(t_2)
+
\theta_i(t_1)\delta\theta_i(t_2)
\right].
\label{spinbdryvar}
\end{equation}
Using the Grassmann identity
\begin{equation}
\delta\theta_i(t_1)\theta_i(t_2)
=
-\theta_i(t_2)\delta\theta_i(t_1),
\label{grassidentity}
\end{equation}
the total boundary contribution takes the form
\begin{equation}
\left.\delta S\right|_{\partial}
=
\frac{i}{2}
\left[
\theta_i(t_1)+\theta_i(t_2)
\right]
\left[
\delta\theta_i(t_2)-\delta\theta_i(t_1)
\right].
\label{spintotalvar}
\end{equation}
The boundary contribution thus vanishes for the
antiperiodic condition (\ref{antiperiodic}).

This example illustrates the distinction between the
information contained in local first-order equations and
that associated with the physical nature of their boundary
conditions. The same distinction motivates our formulation
of the renormalization group. While the beta functions and
anomalous dimensions determine the local RG evolution,
the boundary data characterize the physical information
with which this evolution is supplemented. In the infrared
sector of QCD, where collective excitations and geometric
structures emerge, the Hamiltonian formulation provides a
natural setting in which to investigate the interplay
between these two aspects of the renormalization group.


\section{The renormalization group as a Hamiltonian problem}

The preceding example motivates a variational formulation of the RG
in which the endpoint data are specified explicitly.

Using the RG parameter
$t=\ln\mu$ introduced above, the renormalization-group equation for
a set of couplings $g^a$ can be written as
\begin{equation}
\dot g^a=\beta^a(g).
\label{rgeq}
\end{equation}
The scaling of renormalized quantities $F^A(g,\mu)$ is governed
by the Callan--Symanzik equation

\begin{equation}
\left[
\mu\frac{\partial}{\partial\mu}
+
\beta^a(g)\frac{\partial}{\partial g^a}
\right]F^A
=
-\Gamma^A_{\phantom{A}B}(g)F^B ,
\label{cseq}
\end{equation}
where $\Gamma^A_{\phantom{A}B}$ is the anomalous-dimension matrix.

Along a renormalization-group trajectory,
\begin{equation}
\frac{d}{dt}
=
\mu\frac{\partial}{\partial\mu}
+
\beta^a(g)\frac{\partial}{\partial g^a},
\label{rgtrajectory}
\end{equation}
so that the Callan--Symanzik equation takes the form
\begin{equation}
\dot F^A
=
-\Gamma^A_{\phantom{A}B}(g)F^B.
\label{csflow}
\end{equation}
The beta-function contribution is included in the total derivative
\eqref{rgtrajectory}.

Equations~\eqref{rgeq} and \eqref{csflow} can be incorporated
simultaneously into a first-order action by introducing conjugate variables
$p_a$ and $\Pi_A$,
\begin{equation}
S_{\rm RG}
=
\int_{t_{\rm UV}}^{t_{\rm IR}}dt
\left\{
p_a\left[\dot g^a-\beta^a(g)\right]
+
\Pi_A
\left[
\dot F^A+
\Gamma^A_{\phantom{A}B}(g)F^B
\right]
\right\}
+
B_{\rm UV-IR}.
\label{rgaction}
\end{equation}
Variation with respect to $p_a$ reproduces the usual
renormalization-group equation,
\begin{equation}
\dot g^a=\beta^a(g),
\label{rgeom}
\end{equation}
whereas variation with respect to $\Pi_A$ reproduces the
Callan--Symanzik equation along the trajectory,
\begin{equation}
\dot F^A
=
-\Gamma^A_{\phantom{A}B}(g)F^B.
\label{cseom}
\end{equation}
Hamiltonian formulations of RG evolution have been considered
previously \cite{Dolan1995, Nair}. Our focus is on endpoint polarizations
and their application to collective infrared dynamics.

The corresponding Hamiltonian is
\begin{equation}
H_{\rm RG}(g,F;p,\Pi)
=
p_a\beta^a(g)
-
\Pi_A\Gamma^A_{\phantom{A}B}(g)F^B.
\label{rgham}
\end{equation}
The remaining equations obtained by varying with respect to $g^a$ and
$F^A$ are
\begin{equation}
\dot p_a
=
-p_b\frac{\partial\beta^b}{\partial g^a}
+
\Pi_A
\frac{\partial\Gamma^A_{\phantom{A}B}}{\partial g^a}
F^B ,
\label{peom}
\end{equation}
and
\begin{equation}
\dot\Pi_A
=
\Pi_B\Gamma^B_{\phantom{B}A}(g).
\label{pieom}
\end{equation}
Therefore, the Callan--Symanzik equations can be regarded as a
Hamiltonian system on the enlarged phase space
\begin{equation}
(g^a,F^A;p_a,\Pi_A).
\label{extendedphase}
\end{equation}
The canonical variables are independent, so derivatives with respect
to $g^a$ are taken at fixed $F^A$ and $\Pi_A$:

\begin{equation}
\left.\frac{\partial F^A}{\partial g^a}\right|_{\rm phase\ space}=0,
\qquad
\left.\frac{\partial \Pi_A}{\partial g^a}\right|_{\rm phase\ space}=0.
\end{equation}
Their dependence along a particular RG trajectory follows from
the equations of motion.

On shell, the variation of \eqref{rgaction} reduces to the
endpoint contribution

\begin{equation}
\left.\delta S_{\rm RG}\right|_{\partial}
=
p_{a,\rm IR}\delta g^a_{\rm IR}
-
p_{a,\rm UV}\delta g^a_{\rm UV}
+
\Pi_{A,\rm IR}\delta F^A_{\rm IR}
-
\Pi_{A,\rm UV}\delta F^A_{\rm UV}
+
\delta B_{\rm UV-IR}.
\label{rgboundvar}
\end{equation}
The local RG vector field,

\begin{equation}
\dot g^a=\beta^a(g),
\qquad
\dot F^A=-\Gamma^A_{\phantom{A}B}(g)F^B,
\label{extendedflow}
\end{equation}
does not specify which endpoint variables must be held fixed.
The boundary term must be compatible with the prescribed data
so that

\begin{equation}
\delta S_{\rm RG}=0.
\label{stationaryrg}
\end{equation}
For each canonical pair, the endpoint polarization may be chosen
by fixing either the coordinate or its conjugate momentum.
The two sets of variables are

\begin{equation}
(g^a,F^A)
\qquad\hbox{and}\qquad
(p_a,\Pi_A)
\label{canonicalsets}
\end{equation}
may be chosen at the UV and IR endpoints.

We introduce the collective notation

\begin{equation}
Q^I=(g^a,F^A),
\qquad
P_I=(p_a,\Pi_A).
\label{collectiveQP}
\end{equation}
The action can then be written in Hamiltonian form as
\begin{equation}
S_{\rm bulk}
=
\int_{t_{\rm UV}}^{t_{\rm IR}}dt
\left[
P_I\dot Q^I-H_{\rm RG}(Q,P)
\right].
\label{hamaction}
\end{equation}
On the equations of motion, its variation reduces to
\begin{equation}
\left.\delta S_{\rm bulk}\right|_{\rm on-shell}
=
P_{I,\rm IR}\delta Q^I_{\rm IR}
-
P_{I,\rm UV}\delta Q^I_{\rm UV}.
\label{onshellvar}
\end{equation}
This expression directly determines the boundary term required for each
choice of fixed endpoint data.

If the coordinates $Q^I$ are held fixed at both endpoints,
\begin{equation}
\delta Q^I_{\rm UV}
=
\delta Q^I_{\rm IR}
=
0,
\label{QQcondition}
\end{equation}
no Legendre term needs to be added,
\begin{equation}
B_{\rm bdry}^{(QQ)}=0.
\label{QQboundary}
\end{equation}
If $Q^I_{\rm UV}$ and $P_{I,\rm IR}$ are held fixed, one requires
\begin{equation}
B_{\rm bdry}^{(QP)}
=
-P_{I,\rm IR}Q^I_{\rm IR}.
\label{QPboundary}
\end{equation}
Indeed,
\begin{equation}
\left.
\delta
\left(
S_{\rm bulk}+B_{\rm bdry}^{(QP)}
\right)
\right|_{\rm on-shell}
=
-P_{I,\rm UV}\delta Q^I_{\rm UV}
-
Q^I_{\rm IR}\delta P_{I,\rm IR},
\label{QPvariation}
\end{equation}
which vanishes when
\begin{equation}
\delta Q^I_{\rm UV}=0,
\qquad
\delta P_{I,\rm IR}=0.
\label{QPcondition}
\end{equation}
Similarly, for
\begin{equation}
\delta P_{I,\rm UV}=0,
\qquad
\delta Q^I_{\rm IR}=0,
\label{PQcondition}
\end{equation}
the appropriate term is
\begin{equation}
B_{\rm bdry}^{(PQ)}
=
P_{I,\rm UV}Q^I_{\rm UV},
\label{PQboundary}
\end{equation}
whereas if the momenta are held fixed at both endpoints,
\begin{equation}
\delta P_{I,\rm UV}
=
\delta P_{I,\rm IR}
=
0,
\label{PPcondition}
\end{equation}
one requires
\begin{equation}
B_{\rm bdry}^{(PP)}
=
P_{I,\rm UV}Q^I_{\rm UV}
-
P_{I,\rm IR}Q^I_{\rm IR}.
\label{PPboundary}
\end{equation}
We thus have the four natural canonical choices
\begin{equation}
(Q_{\rm UV},Q_{\rm IR}),
\qquad
(Q_{\rm UV},P_{\rm IR}),
\qquad
(P_{\rm UV},Q_{\rm IR}),
\qquad
(P_{\rm UV},P_{\rm IR}),
\label{fourchoices}
\end{equation}
with the corresponding surface terms
\begin{equation}
0,
\qquad
-P_{\rm IR}Q_{\rm IR},
\qquad
P_{\rm UV}Q_{\rm UV},
\qquad
P_{\rm UV}Q_{\rm UV}-P_{\rm IR}Q_{\rm IR}.
\label{fourboundaryterms}
\end{equation}
These choices correspond to different endpoint polarizations of
the same Hamiltonian system. To distinguish their surface terms
from the generating function, consider the canonical map

\begin{equation}
{\cal M}_{\rm RG}:
\left(Q^I_{\rm UV},P_{I,\rm UV}\right)
\longrightarrow
\left(Q^I_{\rm IR},P_{I,\rm IR}\right).
\label{RGmap}
\end{equation}
A generating function exists in a given polarization only if
the corresponding endpoint variables provide independent
coordinates on the canonical map. Here $Q^I_{\rm IR}$ is
determined entirely by $Q^I_{\rm UV}$. Moreover, for the
first-order action

\begin{equation}
S_{\rm bulk}
=
\int_{t_{\rm UV}}^{t_{\rm IR}}dt\,
P_I\left[\dot Q^I-V^I(Q)\right],
\label{constraintaction}
\end{equation}
the equations of motion imply
\begin{equation}
\dot Q^I=V^I(Q),
\label{constraintflow}
\end{equation}
and therefore
\begin{equation}
\left.S_{\rm bulk}\right|_{\rm on-shell}=0.
\label{constraintbulkonzero}
\end{equation}
The type--1 representation $W_1(Q_{\rm UV},Q_{\rm IR})$
is therefore singular. By contrast, the mixed polarization
$(Q_{\rm UV},P_{\rm IR})$ provides independent endpoint
data, and its generating function follows from the action

\begin{equation}
S^{(QP)}
=
S_{\rm bulk}
-
P_{I,\rm IR}Q^I_{\rm IR}.
\label{QPaction}
\end{equation}
Evaluated on a classical RG trajectory, this gives
\begin{equation}
W_{QP}(Q_{\rm UV},P_{\rm IR})
=
\left.S^{(QP)}\right|_{\rm on-shell}
=
-
P_{I,\rm IR}Q^I_{\rm IR}
(Q_{\rm UV},P_{\rm IR}),
\label{WQP}
\end{equation}
where $Q^I_{\rm IR}(Q_{\rm UV},P_{\rm IR})$ is determined by the
canonical RG map. The corresponding canonical relations are
\begin{equation}
P_{I,\rm UV}
=
-\frac{\partial W_{QP}}
{\partial Q^I_{\rm UV}},
\qquad
Q^I_{\rm IR}
=
-\frac{\partial W_{QP}}
{\partial P_{I,\rm IR}}.
\label{WQPrelations}
\end{equation}
Other generating functions require the corresponding endpoint
projection to be nondegenerate. Variational admissibility of a
polarization does not guarantee the existence of its generating
function or its physical suitability. In QCD, the choice of
endpoint data must account for the distinct physical descriptions
of the ultraviolet and infrared regimes, particularly the emergence
of collective degrees of freedom in the latter.

The extended phase space also makes explicit the persistence
of anomalous scaling at a fixed point. If

\begin{equation}
\beta^a(g_\ast)=0,
\label{betafixed}
\end{equation}
it may simultaneously happen that
\begin{equation}
\Gamma^A_{\phantom{A}B}(g_\ast)\neq0.
\label{gammafixed}
\end{equation}
In that case,
\begin{equation}
\dot g^a=0,
\qquad
\dot F^A
=
-\Gamma^A_{\phantom{A}B}(g_\ast)F^B.
\label{fixedtransport}
\end{equation}
The flow of the couplings stops, but a nontrivial transport of the
renormalized quantities remains. This distinction will be central when
considering the infrared sector of QCD.

\section{Ultraviolet Perturbative Examples}

Two elementary examples illustrate the canonical transport
associated with conventional perturbative RG evolution.

Consider a single coupling $g$ and the action
\begin{equation}
S_{\rm RG}
=
\int_{t_{\rm UV}}^{t_{\rm IR}}
dt\,p\left[\dot g-\beta(g)\right],
\qquad
t=\ln\mu.
\label{singleRGaction}
\end{equation}
The equations of motion obtained through independent variations of $p$
and $g$ are
\begin{equation}
\dot g=\beta(g),
\label{singleflow}
\end{equation}
and
\begin{equation}
\dot p=-p\,\beta'(g).
\label{singlepflow}
\end{equation}
The first equation is precisely the usual renormalization-group equation,
\begin{equation}
\mu\frac{dg}{d\mu}=\beta(g),
\label{usualRG}
\end{equation}
whereas the second determines the evolution of the canonically conjugate
variable required by the variational formulation.

\subsubsection{Trivial Fixed Point}

The simplest example corresponds to
\begin{equation}
\beta(g)=0.
\label{trivialbeta}
\end{equation}
The equations of motion reduce to
\begin{equation}
\dot g=0,
\qquad
\dot p=0,
\label{trivialflow}
\end{equation}
and therefore
\begin{equation}
g_{\rm IR}=g_{\rm UV},
\qquad
p_{\rm IR}=p_{\rm UV}.
\label{identitymap}
\end{equation}
The canonical map is the identity in this case.

\subsubsection{Perturbative Flow}

Let us now consider the simple case
\begin{equation}
\beta(g)=bg^2,
\label{betabg2}
\end{equation}
where $b$ is a constant. The renormalization-group equation is
\begin{equation}
\dot g=bg^2.
\label{bgflow}
\end{equation}
Integrating between $t_{\rm UV}$ and $t_{\rm IR}$ we obtain
\begin{equation}
\frac{1}{g_{\rm IR}}
-
\frac{1}{g_{\rm UV}}
=
-b(t_{\rm IR}-t_{\rm UV}),
\label{integratedflow}
\end{equation}
or, equivalently,
\begin{equation}
\frac{1}{g_{\rm IR}}
=
\frac{1}{g_{\rm UV}}
+
b\ln\frac{\mu_{\rm UV}}{\mu_{\rm IR}}.
\label{runningg}
\end{equation}

The equation for the conjugate momentum is
\begin{equation}
\dot p=-2bgp.
\label{ppertflow}
\end{equation}
Using $\dot g=bg^2$, we may eliminate $t$ and write
\begin{equation}
\frac{dp}{dg}
=
-\frac{2p}{g}.
\label{dpdg}
\end{equation}
Its integration gives
\begin{equation}
pg^2=C,
\label{pgconstant}
\end{equation}
where $C$ is constant along the flow. In particular,
\begin{equation}
p_{\rm IR}g_{\rm IR}^2
=
p_{\rm UV}g_{\rm UV}^2.
\label{pmatching}
\end{equation}
The RG map in phase space is then completely determined by
\begin{equation}
(g_{\rm UV},p_{\rm UV})
\longrightarrow
(g_{\rm IR},p_{\rm IR}),
\label{pertmap}
\end{equation}
where $g_{\rm IR}$ is given by \eqref{runningg} and $p_{\rm IR}$ by
\eqref{pmatching}.

The projection of this canonical map onto coupling space
recovers the conventional flow,

\begin{equation}
\mu\frac{dg}{d\mu}=\beta(g).
\label{projectedflow}
\end{equation}
In particular, if the value of the coupling is specified at a reference
scale,
\begin{equation}
g(\mu_{\rm UV})=g_{\rm UV},
\label{UVdatum}
\end{equation}
the value of $g$ at any other scale is determined by the flow equation.
The values $g_{\rm UV}$ and $g_{\rm IR}$ therefore do not constitute
independent data: they must be related by the renormalization-group
trajectory.

We now turn to the Born--Oppenheimer--Berry (BOB) description
\cite{Berry1984} of infrared QCD, where collective degrees of
freedom provide a different physical setting for the endpoint data.


\section{The Variational Problem in the Infrared Regime of QCD}
\label{sec:IRvariational}

The deep infrared regime of QCD differs from that of QED because
of the nonlinear dynamics of the gauge fields. The Born--Oppenheimer--Berry
(BOB) description of this regime \cite{breslau} identifies two coupled
collective dressing channels, quark--gluon and gluon--gluon, characterized
by a geometric Berry contribution and a corresponding superselection
structure.

The infrared scale $\Lambda$ originates in the treatment of soft
radiation and dressed states \cite{Bloch:1937pw,Chung:1965zza,kibble1,kibble2,kibble3,KF}.
It is distinct from the Wilsonian scale $\mu$. Independence of inclusive
observables from an auxiliary infrared separation scale motivates an RG
description of the dressing functions. In the effective regime considered
here, a stationary coupling coexists with collective scaling and channel
mixing. We investigate how this structure determines the canonical RG
transport and the long-wavelength excitation spectrum.

The physical Hamiltonian generates time evolution of the collective
configurations, whereas the RG Hamiltonian generates evolution in
$\tau=\ln\Lambda$. The dressing functions entering the excitation
energies connect these two descriptions.

\subsection{From the Berry-dressed Hamiltonian to RG variables}

The BOB construction provides a concrete realization of infrared collective dynamics in which the canonical structure is modified by an emergent geometric connection. We first explain the scale covariance of the associated magnetic algebra and then identify the variables relevant to the Hamiltonian RG formulation.

The gauge field is separated into slow and fast components, $A_\mu=A_{\mu,<}+A_{\mu,>}$. At fixed slow configuration $A^I$, the fast-sector Hamiltonian possesses instantaneous states $|n;A\rangle$. Adiabatic projection onto a retained subspace produces the non-abelian Berry connection \cite{Berry1984,WilczekZee1984}
\begin{equation}
(\mathcal A_I)_{mn}[A]=i\left\langle m;A\left|\frac{\delta}{\delta A^I}\right|n;A\right\rangle.
\end{equation}
For a multidimensional retained subspace, the connection is matrix valued. The effective action contains the geometric transport term
\begin{equation}
S_{\mathrm B}=\int dt\,\mathcal A_I[A]\dot A^I=\int_{\mathcal C}\mathcal A_I[A] \,dA^I,
\end{equation}
with matrix-valued parallel transport understood where appropriate. In a single-state sector, $S_{\mathrm{eff}}[A]=S_{\mathrm{YM}}[A]+S_{\mathrm B}$.

The canonical momentum and Berry-covariant electric field are
\begin{equation}
\Pi_I^{\mathrm{phys}}=E_I+\mathcal A_I[A],\qquad E_I=\Pi_I^{\mathrm{phys}}-\mathcal A_I[A]. \label{momen12}
\end{equation}
The Berry curvature is
\begin{equation}
\mathcal F_{IJ}=\frac{\delta\mathcal A_J}{\delta A^I}-\frac{\delta\mathcal A_I}{\delta A^J}-i[\mathcal A_I,\mathcal A_J],
\end{equation}
and the covariant electric fields obey
\begin{equation}
[A^I,E_J]=i\delta^I{}_J,\qquad [E_I,E_J]=i\mathcal F_{IJ}.
\end{equation}

The scaling content of this algebra can be seen from its mechanical analogue. For
\begin{equation}
L=\frac12\dot x_i\dot x_i+g\,a_i(x)\dot x_i,\qquad \pi_i=p_i-g\,a_i(x),
\end{equation}
one has
\begin{equation}
[x_i,\pi_j]=i\delta_{ij},\qquad [\pi_i,\pi_j]=ig\,b_{ij},\qquad b_{ij}=\partial_i a_j-\partial_j a_i.
\end{equation}
Under $x_i\mapsto\lambda x_i$, covariance of the first commutator requires $\pi_i\mapsto\lambda^{-1}\pi_i$. Hence the second commutator has weight $-2$. In the scale-covariant magnetic sector, the curvature has this weight, $b_{ij}\mapsto\lambda^{-2}b_{ij}$, while $g$ is unchanged. Equivalently, $a_i\mapsto\lambda^{-1}a_i$ (up to gauge), so that
\begin{equation}
\int_{\mathcal C}g\,a_i\,dx_i\mapsto\int_{\mathcal C}g\,a_i\,dx_i.
\end{equation}
With $t\mapsto\lambda^2t$, the kinetic action is invariant as well. Thus the magnetic algebra and the geometric term exhibit the same scale covariance without requiring a rescaling of the coupling.

The BOB algebra has the corresponding structure in collective configuration space. Under the homogeneous transformations
\begin{equation}
A^I\mapsto\lambda A^I,\qquad E_I\mapsto\lambda^{-1}E_I,
\end{equation}
the curvature scales as $\mathcal F_{IJ}\mapsto\lambda^{-2}\mathcal F_{IJ}$ and the connection as $\mathcal A_I\mapsto\lambda^{-1}\mathcal A_I$ up to a geometric gauge transformation. Consequently, the Berry line integral $\int_{\mathcal C}\mathcal A_I\,dA^I$ is invariant. This supplies the algebraic motivation for considering a scale-covariant infrared effective sector with a stationary coupling. 

The emergent Berry connection does not constitute an independent
propagating gauge field and introduces no independent gauge coupling.
The covariance of its algebra motivates an infrared fixed-point regime.
If this covariance is realized as an exact symmetry of the renormalized
sector with a scale-independent effective coupling, then
\begin{equation}
\beta_g^{\mathrm{IR}}=\Lambda \frac{dg}{d\Lambda}=0.
\end{equation}
This is a property of the effective infrared regime, distinct from
the Wilsonian running of microscopic QCD. The geometric scaling of
the collective excitations can nevertheless remain nontrivial.

The retained transverse collective Hamiltonian is, in the
quadratic approximation,\footnote{The quadratic approximation
refers only to the explicit dependence of the effective
Hamiltonian on the transverse gauge and electric fields.
It does not imply a free-field approximation, since the
electric fields satisfy noncanonical commutation relations
determined by the Berry curvature. These relations retain
nontrivial geometric information about the underlying
non-Abelian dynamics. Consequently, the collective infrared
excitations can exhibit nontrivial dynamics and modified
dispersion relations even when the effective Hamiltonian
is quadratic.}
\begin{equation}
H_{\mathrm{eff}}=\frac12\int\frac{d^3k}{(2\pi)^3}\left[E_i^T(-k)E_i^T(k)+|k|^2 A_i^T(-k)A_i^T(k)\right],
\end{equation}
where the electric fields satisfy the Berry-modified algebra. Its dressed normal-mode form is
\begin{equation}
H_{\mathrm{eff}}^{\mathrm{IR}}=\int\frac{d^3k}{(2\pi)^3}\sum_s\epsilon_s(k)N_s(k),\qquad \epsilon_s(k)=\hbar|k|F_s(k,\Lambda).
\end{equation}
The functions $F_s$ describe the geometric dressing of collective modes, with $k$ the physical momentum and $\Lambda$ the renormalization scale.

For the two coupled quark-gluon and gluon-gluon channels, introduce
\begin{equation}
\mathbf F=\begin{pmatrix}F^{(qg)}\\F^{(gg)}\end{pmatrix},\qquad
\Gamma=\begin{pmatrix}\gamma_F&\eta_{FG}\\\eta_{GF}&\gamma_G\end{pmatrix}.
\end{equation}
Their infrared transport is
\begin{equation}
\left(\Lambda\frac{\partial}{\partial\Lambda}-\Gamma\right)\mathbf F(k,\Lambda)=0, \label{IVbranches}
\end{equation}
with $k$ fixed. The matrix $\Gamma$ governs geometric scaling and channel mixing, rather than necessarily coinciding with the perturbative anomalous-dimension matrix of elementary QCD fields. Its eigenvalues are
\begin{equation}
\gamma_\pm=\frac{\gamma_F+\gamma_G}{2}\pm\frac12\sqrt{(\gamma_F-\gamma_G)^2+4\eta_{FG}\eta_{GF}}.
\end{equation}
For constant, diagonalizable $\Gamma$ with real eigenvalues, the eigenbranches satisfy
\begin{equation}
\frac{dF_\alpha}{d\tau}=\gamma_\alpha F_\alpha,\qquad \tau=\ln\Lambda,\qquad \alpha=\pm.
\end{equation}
Using the normalization $F_\alpha(k,k)=1$, one obtains
\begin{equation}
F_\alpha(k,\Lambda)=\left(\frac{|k|}{\Lambda}\right)^{-\gamma_\alpha},\qquad \epsilon_\alpha(k)=\hbar \Lambda^{\gamma_\alpha}|k|^{1-\gamma_\alpha}. \label{definition}
\end{equation}
Thus the effective infrared sector exhibits two distinct properties:
\begin{equation}
\beta_g^{\mathrm{IR}}=0,\qquad \Lambda \frac{\partial\mathbf F}{\partial\Lambda}=\Gamma\mathbf F
\end{equation}
The first describes the stationary effective coupling; the second describes nontrivial scale transport of the Berry-dressed excitations. These dressing equations provide the RG vector field to be represented by the first-order Hamiltonian action below.

\subsection{Hamiltonian RG evolution of the dressed branches}

Recall the extended phase-space action of Sec.~II, with coordinates $(g^a,F^A)$ and conjugate variables $(p_a,\Pi_A^{\mathrm{RG}})$. In the convention of Eq.~(\ref{IVbranches}), the anomalous-dimension matrix entering the general Callan--Symanzik equation of Sec.~II is $\Gamma_{\mathrm{CS}}=-\Gamma$. Thus, for the two eigenbranches and the gauge coupling, the corresponding first-order action is
\begin{equation}
 S_{\mathrm{RG}}=\int_{\tau_{\mathrm{UV}}}^{\tau_{\mathrm{IR}}}d\tau\left\{p_g(\dot g-\beta_g)+\sum_{\alpha=\pm}\Pi_\alpha^{\mathrm{RG}}(\dot F_\alpha-\gamma_\alpha F_\alpha)\right\}+B_{\mathrm{UV-IR}}.
 \label{eq:IVRGaction}
\end{equation}
The bulk Hamiltonian is
\begin{equation}
 H_{\mathrm{RG}}=p_g\beta_g+\sum_{\alpha=\pm}\gamma_\alpha\Pi_\alpha^{\mathrm{RG}}F_\alpha.
 \label{eq:IVHRG}
\end{equation}
Variation with respect to $p_g$ and $\Pi_\alpha^{\mathrm{RG}}$ gives $\dot g=\beta_g$ and $\dot F_\alpha=\gamma_\alpha F_\alpha$. The conjugate equations, for constant $\gamma_\alpha$ in the fixed-point regime, give $\dot\Pi_\alpha^{\mathrm{RG}}=-\gamma_\alpha\Pi_\alpha^{\mathrm{RG}}$. Hence
\begin{equation}
 \frac{d}{d\tau}(\Pi_\alpha^{\mathrm{RG}}F_\alpha)=0.
 \label{eq:IVinvariant}
\end{equation}
For positive branch amplitudes, define
\begin{equation}
 q_\alpha=\ln F_\alpha,\qquad p_\alpha=\Pi_\alpha^{\mathrm{RG}}F_\alpha.
 \label{eq:IVqp}
\end{equation}
This is a canonical transformation because $\Pi_\alpha^{\mathrm{RG}}dF_\alpha=p_\alpha dq_\alpha$. The collective part of the RG action and its Hamiltonian become
\begin{equation}
 S_{\mathrm{IR}}^{\mathrm{bulk}}=\sum_{\alpha=\pm}\int_{\tau_{\mathrm{UV}}}^{\tau_{\mathrm{IR}}}d\tau\,p_\alpha(\dot q_\alpha-\gamma_\alpha),\qquad
 H_{\mathrm{RG}}^{\mathrm{IR}}=\sum_{\alpha=\pm}\gamma_\alpha p_\alpha.
 \label{eq:IVqaction}
\end{equation}
Hamilton's equations are therefore
\begin{equation}
 \dot q_\alpha=\gamma_\alpha,\qquad\dot p_\alpha=0.
 \label{eq:IVqeqs}
\end{equation}
This is the central result of the application. The gauge coupling is stationary in the chosen IR sector, but the logarithmic coordinates of the dressed branches continue to evolve canonically. Their velocities are the geometric scaling exponents $\gamma_\alpha$.

Here the labels UV and IR denote the upper and lower endpoints of the infrared effective scale interval, not a matching to microscopic ultraviolet QCD. Let $T=\tau_{\mathrm{UV}}-\tau_{\mathrm{IR}}=\ln(\Lambda_{\mathrm{UV}}/\Lambda_{\mathrm{IR}})>0$. Integration gives the UV--IR canonical map
\begin{equation}
 q_{\alpha,\mathrm{IR}}=q_{\alpha,\mathrm{UV}}-\gamma_\alpha T,\qquad
 p_{\alpha,\mathrm{IR}}=p_{\alpha,\mathrm{UV}},
 \label{eq:IVmap}
\end{equation}
which is equivalent to
\begin{equation}
 \frac{F_{\alpha,\mathrm{IR}}}{F_{\alpha,\mathrm{UV}}}=\left(\frac{\Lambda_{\mathrm{IR}}}{\Lambda_{\mathrm{UV}}}\right)^{\gamma_\alpha}.
 \label{eq:IVratio}
\end{equation}
The flow is canonical even though $\beta_g^{\mathrm{IR}}=0$. This is precisely what the extended RG phase space adds to a description formulated only in terms of the running gauge coupling.

\subsection{Endpoint variation and physical infrared data}

The Hamilton equations determine the canonical map, but not which endpoint data should be fixed. Varying Eq.~(\ref{eq:IVqaction}) yields
\begin{equation}
 \delta S_{\mathrm{IR}}^{\mathrm{bulk}}=\sum_{\alpha=\pm}\int_{\tau_{\mathrm{UV}}}^{\tau_{\mathrm{IR}}}d\tau\left[(\dot q_\alpha-\gamma_\alpha)\delta p_\alpha-\dot p_\alpha\delta q_\alpha\right]+\sum_{\alpha=\pm}[p_\alpha\delta q_\alpha]_{\tau_{\mathrm{UV}}}^{\tau_{\mathrm{IR}}}.
 \label{eq:IVvariation}
\end{equation}
On shell, only the endpoint variation survives,
\begin{equation}
 \left.\delta S_{\mathrm{IR}}^{\mathrm{bulk}}\right|_{\mathrm{on\ shell}}=\sum_{\alpha=\pm}(p_{\alpha,\mathrm{IR}}\delta q_{\alpha,\mathrm{IR}}-p_{\alpha,\mathrm{UV}}\delta q_{\alpha,\mathrm{UV}}).
 \label{eq:IVendvariation}
\end{equation}
Moreover, $S_{\mathrm{IR}}^{\mathrm{bulk}}=0$ on shell, because $\dot q_\alpha=\gamma_\alpha$. The vanishing bulk action does not mean that the UV--IR map is trivial: Eq.~(\ref{eq:IVmap}) contains its nontrivial transport. The on-shell variational representation of that map must therefore be supplied by an appropriate endpoint functional.

For example, the mixed polarization $\delta q_{\alpha,\mathrm{UV}}=0$ and $\delta p_{\alpha,\mathrm{IR}}=0$ is implemented by
\begin{equation}
 B_{\mathrm{IR}}^{(qp)}=-\sum_{\alpha=\pm}p_{\alpha,\mathrm{IR}}q_{\alpha,\mathrm{IR}}.
 \label{eq:IVLegendre}
\end{equation}
The resulting on-shell variation is
\begin{equation}
 \left.\delta(S_{\mathrm{IR}}^{\mathrm{bulk}}+B_{\mathrm{IR}}^{(qp)})\right|_{\mathrm{on\ shell}}=-\sum_{\alpha=\pm}\left(p_{\alpha,\mathrm{UV}}\delta q_{\alpha,\mathrm{UV}}+q_{\alpha,\mathrm{IR}}\delta p_{\alpha,\mathrm{IR}}\right).
 \label{eq:IVmixed}
\end{equation}
It vanishes for the stated data. This demonstrates the consistency of a boundary polarization; it does not derive that polarization from infrared QCD.

The normalization $F_\alpha(k,k)=1$, equivalently $q_\alpha(k,k)=0$, fixes the integration constant in the scale-transport equation. 
It should not be mistaken for a physical condition at the deep-infrared endpoint. Likewise, $p_\alpha=\Pi_\alpha^{\mathrm{RG}}F_\alpha$ 
is a momentum in RG phase space and is not the physical gauge-field momentum $\Pi_I^{\mathrm{phys}}$ of 
Eq.~(\ref{momen12}). Selecting a physical IR polarization requires additional information connecting the collective Berry states to the RG endpoint data.

\subsection{Infrared spectrum and collective confinement criterion}

The dispersion relation (\ref{definition}) gives, for a collective
excitation of characteristic wavelength $L\sim |k|^{-1}$,
\begin{equation}
 E_\alpha(L)\sim\hbar\Lambda^{\gamma_\alpha}L^{\gamma_\alpha-1}.
 \label{eq:IVconfinement}
\end{equation}
At fixed $\Lambda$, three distinct infrared regimes follow:
\begin{equation}
\lim_{L\to\infty}E_\alpha(L)=
\begin{cases}
0, & \gamma_\alpha<1,\\
\hbar\Lambda, & \gamma_\alpha=1,\\
+\infty, & \gamma_\alpha>1.
\end{cases}
\label{eq:IVspectralregimes}
\end{equation}
Here the normalization of Eq.~(\ref{definition}) is understood,
with $L=|k|^{-1}$. The last regime defines the energetic
confinement criterion for the collective branch: an excitation
of arbitrarily large wavelength requires arbitrarily large energy
within the effective description.

The two-channel structure makes this criterion sensitive not only
to the diagonal anomalous dimensions but also to their mixing.
Define
\begin{equation}
 D=\det(\Gamma-\mathbf 1)
 =(\gamma_F-1)(\gamma_G-1)-\eta_{FG}\eta_{GF},
 \qquad S=\operatorname{tr}\Gamma-2.
 \label{eq:IVspectralinvariants}
\end{equation}
Since $D=(\gamma_+-1)(\gamma_--1)$ and
$S=\gamma_++\gamma_--2$, for real eigenvalues the spectral classification is
\begin{eqnarray}
\gamma_+>1,\quad\gamma_->1
 &\Longleftrightarrow& D>0,\quad S>0,\nonumber\\
\gamma_+<1,\quad\gamma_-<1
 &\Longleftrightarrow& D>0,\quad S<0,\nonumber\\
\gamma_-<1<\gamma_+
 &\Longleftrightarrow& D<0.
\label{eq:IVclassification}
\end{eqnarray}
The critical surface $D=0$ corresponds to at least one branch
with $\gamma_\alpha=1$. These criteria are invariant under a
constant change of channel basis.

In particular, suppose that both diagonal exponents are below
unity, $\gamma_F<1$ and $\gamma_G<1$. In the absence of mixing,
both branches are soft. If the mixing satisfies
\begin{equation}
\eta_{FG}\eta_{GF}>(1-\gamma_F)(1-\gamma_G),
\label{eq:IVmixingthreshold}
\end{equation}
then $D<0$, and one collective eigenbranch becomes energetically
suppressed at long wavelengths while the other remains soft.
Thus channel mixing can change the qualitative infrared spectrum,
even with a stationary effective coupling. Conversely, if both
diagonal exponents exceed unity, sufficiently strong positive
mixing can convert one of the two energetically confined branches
into a soft branch.

A further consequence is the relative scaling of the two energies,
\begin{equation}
\frac{\epsilon_+(k)}{\epsilon_-(k)}=
\left(\frac{\Lambda}{|k|}\right)^{\gamma_+-\gamma_-}.
\label{eq:IVspectralratio}
\end{equation}
For distinct real eigenvalues, the two collective branches become
parametrically separated as $|k|/\Lambda\to0$. Finite positive
constant matching normalizations multiply this ratio by a constant
without changing its scaling exponent.

The infrared classification therefore depends on the spectrum of
the full anomalous-dimension matrix rather than on the beta function
alone or on the separate diagonal channel exponents. The canonical
RG map transports the dressing data, while the eigenvalues of its
flow determine the energetic character of the collective branches.
The criterion applies to the collective spectrum and does not by
itself establish a Wilson-loop area law. Its realization in QCD
depends on the dynamical values of the matrix elements of $\Gamma$.

\section{Generating Function of the Infrared Map}

The Hamiltonian analysis of Sec.~IV establishes the canonical UV--IR map for the two collective Berry-dressed branches.
In particular, Eqs.~(\ref{eq:IVqaction})--(\ref{eq:IVmixed}) determine the Hamiltonian evolution and the endpoint term required by the mixed
polarization.

We now construct the generating function associated with this canonical transformation.

For the mixed polarization $(q_{\alpha,\mathrm{UV}},p_{\alpha,\mathrm{IR}})$, the appropriate endpoint term was obtained in Eq.~(\ref{eq:IVLegendre}).
Since the bulk action vanishes on shell, the generating function is determined entirely by this boundary contribution:
\begin{equation}
W_{qp}(q_{\mathrm{UV}},p_{\mathrm{IR}})
=
-\sum_{\alpha=\pm}
p_{\alpha,\mathrm{IR}}q_{\alpha,\mathrm{IR}}.
\end{equation}

Using the canonical map derived in Eq.~(\ref{eq:IVmap}),
\begin{equation}
q_{\alpha,\mathrm{IR}}
=
q_{\alpha,\mathrm{UV}}-\gamma_\alpha T,
\end{equation}

we obtain
\begin{equation}
W_{qp}(q_{\mathrm{UV}},p_{\mathrm{IR}})
=
-\sum_{\alpha=\pm}
p_{\alpha,\mathrm{IR}}
\left(q_{\alpha,\mathrm{UV}}-\gamma_\alpha T\right).
\end{equation}

Equivalently,
\begin{equation}
W_{qp}
=
-\sum_{\alpha=\pm}
p_{\alpha,\mathrm{IR}}q_{\alpha,\mathrm{UV}}
+
T\sum_{\alpha=\pm}\gamma_\alpha p_{\alpha,\mathrm{IR}}.
\end{equation}

The second term explicitly encodes the geometric infrared
transport over the RG interval $T$.

The canonical relations are

\begin{equation}
p_{\alpha,\mathrm{UV}}
=
-\frac{\partial W_{qp}}
{\partial q_{\alpha,\mathrm{UV}}},
\qquad
q_{\alpha,\mathrm{IR}}
=
-\frac{\partial W_{qp}}
{\partial p_{\alpha,\mathrm{IR}}}.
\end{equation}

Differentiating the generating function gives

\begin{equation}
p_{\alpha,\mathrm{UV}}=p_{\alpha,\mathrm{IR}},
\qquad
q_{\alpha,\mathrm{IR}}
=
q_{\alpha,\mathrm{UV}}-\gamma_\alpha T,
\end{equation}
in agreement with the Hamiltonian evolution obtained
in Sec.~IV.

Thus, although the bulk action vanishes on shell,
the canonical UV--IR transformation remains nontrivial.
Its scale dependence is explicitly encoded in the
endpoint generating function.

The distinction between the perturbative ultraviolet and collective infrared regimes is now transparent. In the ultraviolet, the coupling evolves according to $\dot g=\beta(g)$. In the scale-covariant infrared sector considered here, by contrast,
\begin{equation}
\beta_g^{\mathrm{IR}}=0,
\qquad
\dot q_\alpha=\gamma_\alpha,
\qquad
\gamma_\alpha\neq0\ \text{in general}.
\end{equation}
The gauge coupling is stationary while the Berry-dressed collective variables undergo nontrivial canonical transport. The RG Hamiltonian fixes this bulk transport, whereas the physical characterization of the endpoints determines which boundary data should be prescribed. The endpoint Legendre term makes the variational principle compatible with those data; it does not, by itself, select the physical infrared boundary condition.

\section{Physical consequences and conclusions}

\subsection{Infrared matching and spectral normalization}

The scale equations determine the transport of the collective branches,
but do not determine their absolute normalization. To exhibit this
explicitly, let $C_\alpha(k)=F_\alpha(k,k)>0$. The general solution of
Eq.~(\ref{IVbranches}) in a diagonalizable, constant-$\Gamma$ regime is
\begin{equation}
 F_\alpha(k,\Lambda)=C_\alpha(k)
 \left(\frac{\Lambda}{|k|}\right)^{\gamma_\alpha}.
 \label{eq:generalnorm}
\end{equation}
The choice $C_\alpha=1$ used above is a normalization prescription.
If homogeneous scaling in $k$ and $\Lambda$ is imposed on the
dimensionless dressing functions, $C_\alpha$ is constant in the
absence of other scales. Otherwise its momentum dependence must be
specified by matching. The RG variational principle alone does not
fix these constants.

For a positive, finite constant $C_\alpha$, the physical excitation
energy is
\begin{equation}
 \epsilon_\alpha(k)=\hbar C_\alpha
 \Lambda^{\gamma_\alpha}|k|^{1-\gamma_\alpha}.
\end{equation}
At fixed $\Lambda$, the distinction between $\gamma_\alpha<1$,
$\gamma_\alpha=1$ and $\gamma_\alpha>1$ is therefore independent of
this normalization. The last case describes an energetic suppression
of arbitrarily long-wavelength collective excitations within the
assumed effective description. It is not, by itself, a derivation of
an area law for Wilson loops or of confinement in microscopic QCD.

\subsection{Canonical transport of infrared response data}

The conjugate variables acquire a concrete interpretation when a
physical endpoint functional is specified. Let $\mathcal O_{\rm IR}$
be a differentiable function of the infrared dressing amplitudes at
fixed momentum and fixed external parameters. Adding
$-\mathcal O_{\rm IR}(\mathbf F_{\rm IR})$ to the bulk action and fixing
$\mathbf F$ at the initial endpoint gives the natural final condition
\begin{equation}
 \Pi^{\rm RG}_{\alpha,\rm IR}=
 \frac{\partial\mathcal O_{\rm IR}}{\partial F_{\alpha,\rm IR}}.
 \label{eq:responsebc}
\end{equation}
Here the units of the auxiliary conjugate variables are chosen to
match those of the endpoint functional. For constant $\Gamma$ in the
original channel basis, write
\begin{equation}
 U=\exp[-\Gamma T],\qquad
 \mathbf F_{\rm IR}=U\mathbf F_{\rm UV}.
\end{equation}
The conjugate equation then yields, using column vectors for the
momenta,
\begin{equation}
 \boldsymbol\Pi^{\rm RG}_{\rm UV}
 =U^{\mathsf T}\boldsymbol\Pi^{\rm RG}_{\rm IR}.
 \label{eq:responsemap}
\end{equation}
Consequently, the response to initial dressing data is precisely the
pullback of the final response under the RG map. This follows also
from the chain rule and provides an explicit interpretation of the
auxiliary canonical momenta once an endpoint observable is supplied.

For example, for a single collective excitation with
$\mathcal O_{\rm IR}=\hbar |k|F_{\alpha,\rm IR}$,
Eq.~(\ref{eq:responsebc}) gives
\begin{equation}
 p_{\alpha,\rm IR}=\epsilon_{\alpha,\rm IR},\qquad
 p_{\alpha,\rm UV}=p_{\alpha,\rm IR}.
\end{equation}
This equality is a statement about the response to logarithmic
normalization data, not conservation of the physical excitation
energy under changes of $\Lambda$. The endpoint functional is an
additional input, and the construction does not select a vacuum or
fix the spectral normalization without independent physical data.

\subsection{Conclusions}

In this work, we have developed a Hamiltonian variational
formulation of the renormalization group in which the
conditions at the endpoints of the RG evolution are treated
as an integral part of the physical problem. The distinction
between local scale evolution and boundary data provides
a unified framework for describing the running of couplings,
anomalous scaling, canonical transport and the relation
between different scale regimes. In particular, the
construction of mixed generating functions establishes
an explicit canonical relation between the endpoint data,
even when the bulk RG action vanishes on shell.

The physical relevance of this formulation becomes apparent
in the Born--Oppenheimer--Berry description of infrared QCD.
The geometric Berry contribution modifies the canonical
algebra of the collective gauge degrees of freedom and
provides the basis for an infrared effective theory in
which a stationary gauge coupling coexists with nontrivial
anomalous scaling. The quark--gluon and gluon--gluon dressing
clouds constitute two interacting collective channels whose
scale evolution is governed by an effective
anomalous-dimension matrix.

The eigenvalues of this matrix determine the infrared
dispersion relations,
\begin{equation}
\epsilon_\pm(k)
=
\hbar\Lambda^{\gamma_\pm}|k|^{1-\gamma_\pm},
\end{equation}
and distinguish three qualitatively different regimes:
soft collective excitations, critical modes with finite
long-wavelength energies, and energetically confined
collective excitations. This classification admits a
basis-independent formulation in terms of the trace and
determinant of the anomalous-dimension matrix.

A particularly interesting consequence is that the mixing
between the quark--gluon and gluon--gluon channels can
qualitatively modify the infrared spectrum. Even when both
uncoupled channels exhibit soft behavior, their interaction
can generate an eigenbranch whose excitation energy
diverges in the long-wavelength limit. Within the
scale-covariant effective description, collective
confinement can therefore emerge as a consequence of
channel mixing rather than solely from the anomalous
scaling of the individual channels.

The distinction between the Wilsonian scale $\mu$ and
the infrared dressing scale $\Lambda$ is essential to
this physical picture. Whereas the former characterizes
the conventional renormalization-group evolution of
microscopic QCD, the latter governs the scaling of the
collective infrared excitations. The Hamiltonian
formulation developed here identifies the canonical
structure of this infrared evolution and establishes
how physical boundary data and spectral response
functions enter its variational description.

The determination of the anomalous-dimension matrix
and the matching of the collective infrared sector
to microscopic QCD constitute natural extensions of
the present approach. More generally, our results
suggest that the interplay between emergent geometry,
collective excitations and renormalization-group
boundary conditions provides a complementary
perspective on the nonperturbative infrared structure
of gauge theories.
\section*{Acknowledgments}

This paper is based on three talks delivered by the author,
two at the Department of Physics of the Technical University
of Munich (TUM) and the other at the
{\it Confinement XVII Conference} in Wroclaw, Poland.
The author is grateful to the members of the phenomenology
group at TUM, and in particular to Professor Nora Brambilla,
for their warm hospitality. He also thanks Professor
G. Schierholz for stimulating discussions.
The author gratefully acknowledges the generous support
of the Alexander von Humboldt Foundation.
This work was supported by DICYT (USACH)
under Grant No.~042531GR\_REG.

\end{document}